\documentclass[submission,copyright,creativecommons]{eptcs}
\providecommand{\event}{ML 2014}

\usepackage{graphicx}
\usepackage{xcolor}
\usepackage[fleqn]{amsmath}
\usepackage[hang,flushmargin]{footmisc}
\usepackage{graphics}
\usepackage{stmaryrd}
\usepackage{amsthm}

\newcommand{\langl}{\begin{picture}(4.5,7)
\put(1.1,2.5){\rotatebox{60}{\line(1,0){5.5}}}
\put(1.1,2.5){\rotatebox{300}{\line(1,0){5.5}}}
\end{picture}}
\newcommand{\rangl}{\begin{picture}(4.5,7)
\put(.9,2.5){\rotatebox{120}{\line(1,0){5.5}}}
\put(.9,2.5){\rotatebox{240}{\line(1,0){5.5}}}
\end{picture}}

\newcommand{\rang}{\begin{picture}(5,7)
\put(.1,2.5){\rotatebox{135}{\line(1,0){6.0}}}
\put(.1,2.5){\rotatebox{225}{\line(1,0){6.0}}}
\end{picture}}

\definecolor{cmtclr}{rgb}{0.0,0.6,0.0}
\definecolor{kvdclr}{rgb}{0.0,0.0,0.6}
\definecolor{strclr}{rgb}{0.5,0.1,0.0}
\definecolor{prepclr}{rgb}{0.0,0.0,0.0}

\newcommand{\sem}[1]{\llbracket #1 \rrbracket}
\newcommand{\ignp}{\_\kern-.5ex\_\kern-.5ex\_ }
\newcommand{\kvd}[1]{\textnormal{\textcolor{kvdclr}{\sffamily #1}}}
\newcommand{\str}[1]{\textnormal{\textcolor{strclr}{\ttfamily "#1"}}}
\newcommand{\ident}[1]{\textnormal{\sffamily #1}}
\newcommand{\lident}[1]{\textnormal{\sffamily
  \`{}\hspace{-0.25em}\`{}\hspace{-0.1em}#1\`{}\hspace{-0.25em}\`{}}}

\newcommand{\narrow}[1]{\hspace{-0.75em} #1 \hspace{-1.25em}~}

\newtheorem*{theorem*}{Theorem}

\title{In the Age of Web:\\
  \textnormal{\LARGE Typed Functional-First Programming Revisited}}

\author{Tomas Petricek
\institute{University of Cambridge, UK}
\email{tomas@tomasp.net}
\and
Don Syme
\institute{Microsoft Research Cambridge, UK}
\email{don.syme@microsoft.com}
\and
Zach Bray
\institute{Type Inferred Ltd}
\email{zachbray@gmail.com}
}
\def\titlerunning{In the Age of Web: Typed Functional-First Programming Revisited}
\def\authorrunning{T. Petricek, D. Syme, Z. Bray}
\begin{document}
\maketitle


\begin{abstract}
Most programming languages were designed before the age of web.
This matters because the web changes many assumptions that typed functional language designers take
for granted. For example, programs do not run in a closed world, but must instead interact with
(changing and likely unreliable) services and data sources, communication is often asynchronous
or event-driven, and programs need to interoperate with untyped environments.

In this paper, we present how the F\# language and libraries face the challenges posed by the web.
Technically, this comprises using \emph{type providers} for integration with external information
sources and for integration with untyped programming environments, using \emph{lightweight
meta-programming} for targeting JavaScript and \emph{computation expressions} for writing
asynchronous code.

In this inquiry, the holistic perspective is more important than each of the features in isolation.
We use a practical case study as a starting point and look at how F\# language and libraries
approach the challenges posed by the web. The specific lessons learned are perhaps less interesting
than our attempt to uncover hidden assumptions that no longer hold in the age of web.
\end{abstract}

%
%

\section{Introduction}

Among the ML family of languages, F\# often takes a pragmatic approach and emphasizes ease of use
and the ability to integrate with its execution environments\footnote{Historically, this applied
to the .NET runtime, but the same applies to integrating with JavaScript in the web context.} over
other aspects of language design. If you use the F\# language as ML, you get most of the good
well-known properties of ML\footnote{For example, F\# does not require type annotations when used
as ML, but requires them when used with .NET objects.}. However, F\# leaves enough \emph{holes} that
let you use it \emph{not} as ML. This is the space that we explore in this paper.

This additional flexibility makes it possible to use F\# in ways that break the common assumptions
that are often taken for granted in languages such as ML and Haskell\footnote{In a way, we are trying
to uncover the hidden assumptions of the functional programming \emph{research programme} that are
normally ``\emph{rendered unfalsifiable by the methodological decisions of its protagonists}.''
(Lakatos \cite{philosophy-lakatos}, quoted by Chalmers \cite{philosophy-thing})}. The focus on
the web directs our inquiry and provides an angle for reconsidering such assumptions.

Perhaps the most remarkable assumption is the idea that programs fundamentally operate in a closed
world. Although we have learned how to perform FFI and I/O \cite{haskell-ffi,haskell-imperative},
those are treated as dealing with the ``dirty real world''. For a practical solution, we argue that
we need to go much further -- to the extent that deeper integration with (untyped) JavaScript
libraries and (evolving) services inevitably breaks some of the strict type safety requirements.

We do not claim that the F\# approach is the only possible one. Rather, this paper should be seen
as a programming language experiment \cite{philosophy-pl} or an empirical observation of the
approach used by the F\# community. We aim to provide an intriguing exploration of hidden assumptions
and present what can be achieved using a combination of F\# features. We do so by starting with
a simple, yet real-world problem and then exploring a solution. More specifically, the contributions
of this position paper are:

\begin{itemize}
\item We present a case study (Section~\ref{sec:case}) showing how a combination of numerous
  F\# language features can be used for the development of modern web applications. This
  is not a toy demonstration, but an example of how F\# is used in industry.

\item We discuss how type providers make it possible to access external information sources
  in web applications (Section~\ref{sec:tp-data}) and integration with (untyped) programming
  environments such as the JavaScript ecosystem (Section~\ref{sec:tp-lang}).

\item We show how F\# approaches the problem of compilation to JavaScript using a library
  called FunScript (Section~\ref{sec:js-meta}), outlining important practical concerns such as
  interoperability (Section~\ref{sec:js-lib}) and asynchronous execution (Section~\ref{sec:js-async}).

\item Throughout the paper, we discuss how the age of the web breaks the assumptions commonly
  taken for granted in typed functional programming. We revisit the notion of type safety
  in the context of the web (Section~\ref{sec:tp-relative}) and the notion of fixed language
  semantics (Section~\ref{sec:js-relative}).
\end{itemize}

\noindent
In the first part of the paper (Section~\ref{sec:case}), we present a case study of using
F\# for web development. The rest of the paper (Section~\ref{sec:tp} and \ref{sec:js}) discusses
the arising issues in more depth. The source code and running demo for the case study is available
at: \url{http://funscript.info/samples/worldbank}

%

\section{Case Study: Web-based data analytics}
\label{sec:case}
\begin{figure}[!t]
\hspace{3em} \includegraphics[width=35em]{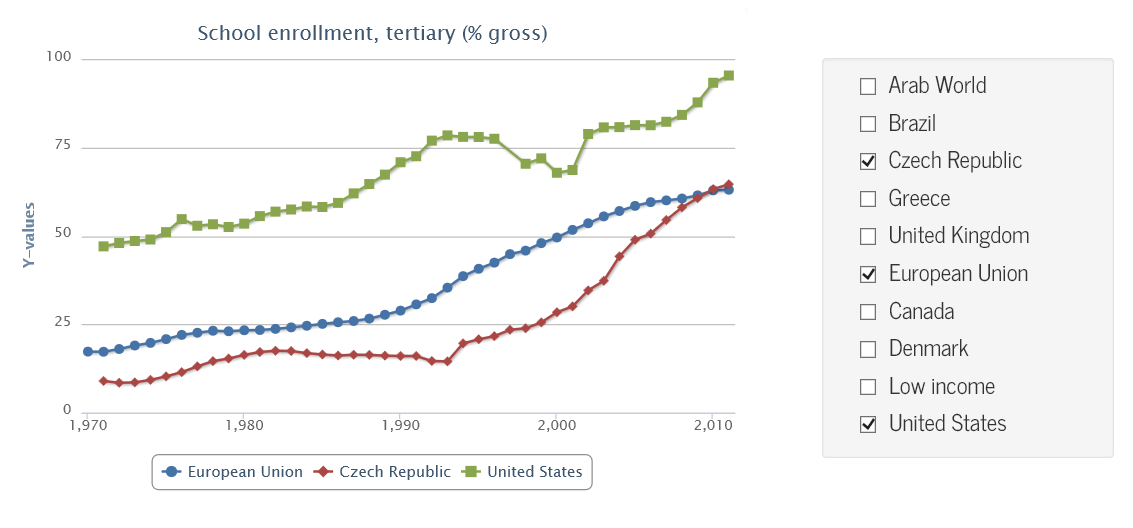}
\caption{Case study -- web application for comparing university enrollment in the world}
\label{fig:wb}
\end{figure}

In this case study, we develop a web application shown in Figure~\ref{fig:wb}, which lets the
user compare university enrollment in a number of selected countries and regions around the world.
The resulting application runs on the client-side (as JavaScript) and fetches data dynamically
from the World Bank \cite{data-wb-schter}.

The application is an example of a web page that could be built in the context of data journalism
\cite{dj-handbook}. As such, it is relatively simple, works with just a single data source and
uses a concrete indicator and a hard-coded list of countries \emph{i.e.}~to illustrate a point
made in an accompanying article.


\subsection{Accessing World Bank data with type providers}

To access the university enrollment information, we first obtain a list of countries using the
World Bank type provider from the F\# Data library \cite{fsharp-data}. The type provider exposes
the individual countries as members of an object (the notation \lident{Country Name} is used for
identifiers with spaces):
\begin{equation*}
\begin{array}{l}
 \kvd{type}~\ident{WorldBank}~=~\ident{WorldBankData}\langl\ident{Asynchronous}=\kvd{true}\rangl
 \\[0.5em]
 \kvd{let}~\ident{data}~=~\ident{WorldBank.GetDataContext}() \\
 \kvd{let}~\ident{countries}~= \\
 \quad~\lbrack~~ \ident{data.Countries.}\lident{European Union} \\
 \qquad   \ident{data.Countries.}\lident{Czech Republic} \\
 \qquad   \ident{data.Countries.}\lident{United Kingdom} \\
 \qquad   \ident{data.Countries.}\lident{United States} ~\rbrack
\end{array}
\end{equation*}
The type provider connects to the World Bank and obtains a list of countries at
\emph{compile-time} and at \emph{edit-time} (when using auto-completion in an editor).
This means that the list is always up-to-date and we get a compile time error when accessing
a country that no longer exists (a property discussed in Section~\ref{sec:tp-data}).

On the first line, we provide a static parameter \ident{Asynchronous}. Static parameters are
resolved at compile-time (or edit-time). Here, we specify that the exposed types for accessing
information should support only non-blocking functions. This is necessary for a web-based
application, because JavaScript only supports non-blocking calls (using callbacks) to fetch the data.


\subsection{Interoperating with JavaScript libraries}
To run the sample application on the client-side we use FunScript \cite{fsharp-funscript}, which
is a library that translates F\# code to JavaScript (Section~\ref{sec:js}). Aside from
running as JavaScript, we also want to use standard JavaScript libraries, including jQuery for DOM
manipulation and Highcharts for charting. FunScript comes with a type provider that imports
TypeScript \cite{ms-typescript} definitions for JavaScript libraries:
\begin{equation*}
\begin{array}{lll}
 \kvd{type}~\ident{j}~&\narrow{=}&~\ident{TypeScript}\langl\str{jquery.d.ts}\rangl \\
 \kvd{type}~\ident{h}~&\narrow{=}&~\ident{TypeScript}\langl\str{highcharts.d.ts}\rangl \\
\end{array}
\end{equation*}
The \textcolor{strclr}{\ttfamily d.ts} files are type annotations created for the TypeScript language.
Here, the type provider mechanism lets us leverage an existing effort for annotating common JavaScript
libraries. The type provider analyses those definitions and maps them into F\# types named \ident{j} and
\ident{h} that contain statically typed functions for calling the JavaScript libraries (we will use
them shortly). The file names are static parameters (same as \ident{Asynchronous} earlier) and are
statically resolved and accessed at compile-time or edit-time.

Importing types for JavaScript libraries into the F\# type system has interesting implications,
because the TypeScript language does not have the traditional type safety property \cite{ms-safets}.
We return to this topic in Section~\ref{sec:tp-lang}. Next, we generate checkboxes that appear on the
right in Figure~\ref{fig:wb}:
\begin{equation*}
\begin{array}{lll}
 \kvd{let}~\ident{jQuery}~\ident{command}~=~\ident{j.jQuery.Invoke}(\ident{command})
 \\[0.5em]
 \kvd{let}~\ident{infos}~=~\ident{countries}~|\rang~\ident{List.map}~(\kvd{fun}~\ident{country}~\rightarrow \\
 \quad \kvd{let}~\ident{inp}~=~\ident{jQuery}(\str{<input>})\ident{.attr}(\str{type}, \str{checkbox}) \\
 \quad \ident{jQuery}(\str{\#panel})\ident{.append}(\ident{inp})\ident{.append}(\ident{country.Name}) \\
 \quad \ident{country.Name}, \ident{country.Indicators}, \ident{el}) \\
\end{array}
\end{equation*}
To manipulate the DOM (Document Object Model), we are using the jQuery library in a way that is
very similar to code that one would write in JavaScript. We define a helper function \ident{jQuery}
(hiding some of the complexities of the mapping) and use it to create the \str{<input>} element and
specify its attributes. Note that members like \ident{append} and \ident{attr} are standard
jQuery patterns. The compiler sees them as ordinary object members. When writing code using F\# editors
based on the F\# Compiler Service \cite{fsharp-fcs}, they also appear in the auto-complete list.

Although the jQuery library is not perfect, it is a de facto standard in web development. The
FunScript type provider makes it possible to integrate with it painlessly without explicitly
specifying any FFI interface and without manual wrapping (see also Section~\ref{sec:js-lib}).

Note that we use a standard F\# function \ident{List.map} to iterate over the countries. The
function passed as an argument has a side-effect of creating the HTML elements, but it also returns
a new list. The result is a list of $\ident{string}\ast\ident{Indicators}\ast\ident{jQuery}$ values
representing the country name, its indicators (for accessing the World Bank data) and the created
DOM object representing the checkbox.


\subsection{Loading data and updating the user interface}
\label{sec:case-loading}

The main part of the sample program is a function \ident{render} that asynchronously fetches
data for selected countries and generates a chart. To keep the code simple, we iterate over the
\ident{infos} list from the previous section and load data for countries one by one:
\begin{equation*}
\begin{array}{lll}
 \kvd{let}~\ident{render}~()~=~\ident{async}~\{ \\
 \quad \kvd{let}~\ident{head}~=~\str{School enrollment, tertiary (\% gross)} \\
 \quad \kvd{let}~\ident{o}~=~\ident{h.HighchartsOptions}() \\
 \quad \ident{o.chart} \leftarrow \ident{h.HighchartsChartOptions}(\ident{renderTo}=\str{plc}) \\
 \quad \ident{o.title} \leftarrow \ident{h.HighchartsTitleOptions}(\ident{text}~=~\ident{head}) \\
 \quad \ident{o.series} \leftarrow \lbrack| ~ |\rbrack
 \\[0.5em]
 \quad \kvd{for}~\ident{name},~\ident{ind},~\ident{check}~\kvd{in}~\ident{infos}~\kvd{do}\\
 \qquad   \kvd{if}~\ident{unbox}\langl \ident{bool} \rangl~(\ident{check.is}(\str{:checked}))~\kvd{then} \\
 \qquad\quad     \kvd{let!}~\ident{v}~=~\ident{ind.}\lident{School enrollment, tertiary (\% gross)} \\
 \qquad\quad     \kvd{let}~\ident{data}~=~\ident{vals}~
                   |\rang~\ident{Seq.map}~(\kvd{fun}~(\ident{k}, \ident{v})~\rightarrow \lbrack|~\ident{number~k};~\ident{number~v} ~|\rbrack)~
                   |\rang~\ident{Array.ofSeq} \\
 \qquad\quad     \ident{opts.series.push}(\ident{h.HighchartsSeriesOptions}(\ident{data}, \ident{name})) ~\}
\end{array}
\end{equation*}
Although the function looks like ordinary code, it is wrapped in the $\ident{async}~\{\ldots\}$
block, which is an F\# computation expression \cite{fsharp-zoo}. The F\# compiler performs
de-sugaring similar to the CPS transformation and interprets keywords such as \kvd{let!} and
\kvd{for} using special operations (monadic bind and others). The \ident{async} identifier
determines that we are writing asynchronous workflow \cite{fsharp-async} that makes it
possible to include non-blocking calls in the block.

Here, the non-blocking call is done when accessing the \lident{School enrollment, tertiary (\% gross)}
indicator using the \kvd{let!} keyword. The indicator is a member (with a name wrapped in
back-ticks to allow spaces) exposed as an asynchronous computation by the World Bank type provider.
The rest of the code is mostly dealing with the DOM and the Highcharts library using the API
imported by FunScript -- we iterate over all checkboxes and generate a new chart series for each
checked country.

Two notable points here are that \ident{async} translated to JavaScript is restricted to a single
thread, which is not the case for ordinary F\# code (Section~\ref{sec:js-async}) and that the
\ident{HighchartOptions} object preserves some of the underlying JavaScript semantics
(Section~\ref{sec:js-lib}). Finally, the last part of the example code registers event handlers that
redraw the chart when the checkbox is clicked:
\begin{equation*}
\begin{array}{lll}
 \kvd{for}~\ignp, \ignp,~\ident{check}~\kvd{in}~\ident{infos}~\kvd{do} \\
 \quad \ident{check.click}(\kvd{fun}~\ignp~\rightarrow \ident{Async.StartImmediate}(\ident{render}()))
\end{array}
\end{equation*}
The \ident{click} operation (exposed by jQuery) takes a function that should be called when the
event occurs. Calling it is a side-effectful operation that registers the handler. As \ident{render}
is an asynchronous operation, we invoke it using the \ident{StartImmediate} primitive from the
F\# library, which starts the computation without waiting for the result (the only way to start
a non-blocking operation in JavaScript).


\subsection{Learning from the case study}

The case study shows that we can develop a simple interactive data visualization (that could
be built, for example, by data journalists) in less than 30 lines of F\# code. The code uses
many typical functional patterns (lists, first-class functions, data types), but also uses features
that are more specific to F\# (type providers, objects, computation expressions).

Before analysing the interesting aspects of this case study, we briefly review the points that we
find appealing and points that many would find unappealing or, at least, peculiar. First, the
appealing points:

\begin{itemize}
\item The ML approach to types and type inference can be extended from (closed-world) data types
  to (open-world) types for rich information sources such as World Bank. The sample code is
  fully statically typed without explicit type annotations. Critically, types are also used for
  exploratory programming when finding indicators using auto-complete in an editor.

\item The case study demonstrates that core ML programming style can be used in the context of
  client-side (JavaScript) web development. We used functional lists, standard higher-order
  functions such as \ident{List.map} in much the same way as when writing ordinary F\#.

\item In addition to standard functional constructs, we were also able to reuse F\# asynchronous
  workflows to write non-blocking code that requests data from a web service (World Bank),
  rather than using error-prone explicit callbacks that are common in JavaScript.

\item Finally, we were able to painlessly call Highcharts and jQuery. No explicit wrapping or
  importing of individual functions and types was necessary. Moreover, despite the differences between
  the F\# and JavaScript object model, the code is close to idiomatic F\#.
\end{itemize}

\noindent
Now, the following list looks at the aspects that appear unappealing or peculiar, especially when
coming from the traditional functional programming background:

\begin{itemize}
\item The World Bank type provider lifts information about countries to the type level. As a result,
  we can easily write $\ident{data.Countries.}\lident{Czech Republic}$, but if Czech Republic is
  removed from the World Bank (and becomes Czechoslovakia again), the code will no longer compile
  (Section~\ref{sec:tp-data}).

\item The TypeScript language is unsound due to covariant generics \cite{ms-safets}. Thus
  importing types from TypeScript definitions introduces a potential unsoundness into the F\#
  code (Section~\ref{sec:tp-lang}).

\item When compiling F\# to JavaScript, the FunScript library does not fully preserve the
  semantics of F\#. For example, numerical types behave as in JavaScript (Section~\ref{sec:js-meta})
  and asynchronous workflows run on a single thread (Section~\ref{sec:js-async}).
\end{itemize}

\noindent
The most notable observation about the above points is that there is often both a positive and
a negative side: we can nicely access World Bank data, but it affects soundness properties;
we can interoperate with JavaScript libraries, but we can not fully hide undesirable JavaScript
behaviours.

The aim of this paper is not to make value judgements and argue what is better. Using the case
study as a basis, we claim that the outlined approach is just \emph{one possible} and that it
\emph{works in practice}. The rest of the paper gives more details about the most important
aspects of the approach and discuss alternatives.

%

\section{Integrating with the open world}
\label{sec:tp}

Type providers \cite{fsharp-typeprov} are a mechanism for integrating external components into a
statically typed programming language. Such components include information sources (such as World
Bank), other environments (here, JavaScript libraries via TypeScript), but type-providers can
also be used for limited meta-programming (the \ident{Asynchronous} parameter can be seen as
a form of meta-programming).


\subsection{How type providers work and fail}
\label{sec:tp-def}

Type providers are libraries that are loaded by the compiler (and editor) and are executed at
compile-time (or edit-time). A type provider builds information about types and makes those
available to the compiler. This is done lazily, so the type provider does not need to provide
types for the entire information space at once. In terms of programming language theory, type
providers change the starting point for a type-checking of a program as follows:
\begin{equation*}
\hspace{8em}
\begin{array}{rclll}
 \emptyset &\narrow{\vdash}& e : \tau
    &\qquad& (\textit{classical, or closed-world})\\
 \pi~(\includegraphics[width=1em,trim=0 1.5em 0 0]{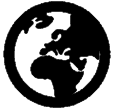})  &\narrow{\vdash}& e : \tau
    &\qquad& (\textit{type providers, or open-world})
\end{array}
\end{equation*}
Rather than starting the type-checking with an empty context, we start type-checking with a
context containing a \emph{projection} of some information from the world. This analogy is useful
for understanding how type providers work and how they can go wrong:

\begin{itemize}
\item \emph{Type provider failure.}
  The projection (a type provider) is implemented in F\# and can fail (for example, if the internet
  is required but unavailable). However, when the type provider succeeds, it generates valid F\#
  code and so unchecked compilation errors cannot happen.

\item \emph{Runtime failure.}
  When the world changes and we \emph{run} the program, we may get a runtime exception. It is
  expected that the provided code will not fail as long as certain assumptions are satisfied (for
  example, countries are not removed from a database). But if the world changes and the assumptions
  no longer hold, the provided code may fail with a runtime exception.

\item \emph{Recompilation failure.}
  When the world changes and we \emph{recompile} the program, the result of the projection can
  differ and so we may not be able to type-check code that used to type-check. This can be seen as
  negative aspect, but there is a clear benefit -- it means we discover an error that would happen
  at runtime earlier during recompilation.
\end{itemize}

\noindent
The above points are \emph{general} points about type providers. In this paper, we look at two
\emph{specific} type providers for World Bank and TypeScript. When discussing type providers, we
look at how the provider (or the projection $\pi$) works and what assumptions it makes
about the world. That is, what world changes cause the provided type to change and what world
changes cause a runtime exception.


\subsection{Integrating with World Bank data}
\label{sec:tp-data}

The World Bank provider \cite{fsharp-data} is a type provider designed specifically for a single
data source. This contrasts with F\# Data type providers for working with CSV, XML and JSON data
that take a sample or schema as a parameter and can thus be used with files of any structure.

The projection implemented by the type provider generates a type called \ident{Countries} that contains
countries as members. Each member returns a value of type \ident{Indicators} that is
also generated and contains all indicators as members. The type provider provides types that are
\emph{erased} during compilation and replaced with runtime implementation. For the World Bank type
provider, the erasure works as follows:
\begin{equation*}
\begin{array}{lcl}
 \sem{~\ident{data.}\lident{Czech Republic}~} &=&
   \ident{data.GetCountries}().\ident{GetCountry}(\str{CZE}) \\
 \sem{~\ident{cz.}\lident{School enrollment, tertiary (\% gross)}~} &=&
   \ident{cz.AsyncGetIndicator}(\str{SE.TER.ENRR})
\end{array}
\end{equation*}
The underlying operations (\ident{GetCountry}, \ident{GetCountries} and \ident{AsyncGetIndicator})
are operations of an underlying runtime library (which is a normal non-provided library). This is
a typical pattern -- type providers generally produce a light layer on top of a rich runtime library.
An important point here is that the type provider uses the \emph{name} of a country or an indicator
for the member name, but at runtime, it uses a \emph{code} of the country or the indicator. This has
interesting implications for the safety properties.

\subsubsection{Safety properties of the type provider}

Given the mapping, we can now look at the assumptions that the World Bank type provider makes about
the world and in which cases it encounters one of the failures defined in Section~\ref{sec:tp-def}.

\begin{itemize}
\item If you are offline when using the type provider (compiling or editing), the type provider will
  not be able to obtain the list of countries or indicators. This is a \emph{type provider failure}.
  The World Bank provider caches the schema, but it still requires internet access for the first time.

\item When a country or indicator is renamed, the compiled code will continue to work (there is no
  \emph{runtime failure}), because the lookup uses the country or the indicator \emph{code}. However,
  recompilation will fail as the member name (which appears in the source code) will be different.

\item Finally, if a country or indicator disappears, we will get \emph{runtime failure} when running
  existing compiled code and \emph{recompilation failure}. Here, the \emph{recompilation
  failure} is a useful indicator that the assumptions made by our code have been violated by
  the data source.
\end{itemize}

\noindent
When contrasted with the traditional view of ML type-safety, the World Bank type provider relaxes
the usual safety conditions. However, this is not the case when compared with standard developer
practices. In particular, if we wrote $\ident{data.GetCountries}().\ident{GetCountry}(\str{CZE})$,
the \emph{runtime failure} behaviour would be the same, but we would lose the useful
\emph{recompilation failure}.

\subsubsection{Discussion of alternatives}

There are two main alternatives worth discussing. First, how would the code look if
we did not use any type provider. Second, what are alternative designs for the World Bank type
provider and how would such alternatives change the safety properties.

\vspace{-1em}
\paragraph{Accessing countries without type providers.} Assuming we want the countries
and regions in Figure~\ref{fig:wb}, we could build a list of country codes and use projection:
\begin{equation*}
\begin{array}{l}
 \kvd{let}~\ident{data}~=~\ident{WorldBank.GetDataContext}() \\
 \kvd{let}~\ident{countryCodes}~=~\lbrack~\str{EU},~\str{CZE},~\str{GBP},~\str{USA}~\rbrack \\
 \kvd{let}~\ident{countries}~=~\ident{countryCodes}~|\rangl~\ident{List.map}~(\ident{data.GetCountries}().\ident{GetCountry}())
\end{array}
\end{equation*}
The motivation for the case study was a data journalism application with a transparent logic
that is easy to understand and modify. The type provider achieves these goals better -- a country
can be added just by looking through an auto-complete list without knowing its code.
As for the safety properties, using explicit country codes has the same runtime properties as using
the type provider. This could be avoided by listing all countries in a given region, but that is
solving a different problem.

In summary, the World Bank type provider fits a specific niche that we were exploring in the case
study. That is, when we want to create an information analysis that accesses specific data from
an information source. With rising popularity of data journalism and Open Government Data
initiatives, we argue that this is an important problem domain.

\vspace{-1em}
\paragraph{Providing safer data accessors.}
The \emph{recompilation failure} when a country is removed from the dataset is an indication that
code would not behave as expected at runtime and so we find it useful. The problematic case is
a \emph{runtime failure} when a country is removed after the program is compiled. One way to
avoid this error would be to change the type of the provided members like
\ident{data.}\lident{Czech Republic} from \ident{Country} to \ident{Country~option}.

However, that is only shifting the burden of error handling from the library to the user. If the
author of the application does not intend to handle errors (for example, by skipping missing data),
they would have to add code that explicitly throws an exception. For the purpose of our case study,
we prefer to keep the end-user code simple -- we can still handle the error using
$\kvd{try}~\ldots~\kvd{with}$.

Whether using option types is more desirable depends on the use case (how we want to handle errors)
and reliability of the data source (countries do not disappear often). The choice can be left to
the user by adding a static parameter (like $\ident{Asynchronous}=\kvd{true}$). For example, the CSV
type provider from F\# Data has a parameter \ident{AssumeMissingValues} that instructs the type
provider to always provide options.

\vspace{-1em}
\paragraph{Accessing data ``as of time''.}
We discussed how to mitigate the problems caused by the fact that the open world changes.
Can we make the world \emph{not} change? This is a sensible question for some data sources
(but not the World Bank). Freebase \cite{google-freebase} supports the notion of ``as of time''.
When calling the API with the \ident{as\_of\_time} parameter it returns data and meta-data that
was present at the specified date.

This would be a desirable option for our case study -- if we want to illustrate point made
by an article, the accompanying application could use data available at the time when the
article was written. However, this feature relies on the ability of the data source -- not all
services have this option.


\subsection{Integrating with JavaScript libraries}
\label{sec:tp-lang}

Type providers are often described as a technique that simplifies data access. The TypeScript
type provider shows that there is a broader range of uses, including integration between different
programming languages. Another example of type provider for language integration is the R
provider \cite{fsharp-rprovider} which imports packages and functions of the statistical
environment R.

The projection implemented by the TypeScript type provider imports type annotations for JavaScript
libraries sa specified by TypeScript \cite{ms-typescript}. The following is an excerpt from the
\str{jquery.d.ts} file that describes the \ident{jQuery} value and the \ident{attr} method used
in our example:
\begin{equation*}
\begin{array}{l}
 \kvd{declare var}~\ident{jQuery}~:~\ident{JQueryStatic};
\\[0.5em]
 \kvd{interface}~\ident{JQueryStatic}~\{ \\
 \quad (\ident{selector}: \kvd{string}, \ident{context?}: \kvd{any}):~\ident{JQuery}; \\
\}
\\
\kvd{interface}~\ident{JQuery}~\{ \\
\quad \ident{attr}(\ident{attributeName}~:~\kvd{string}):~\kvd{string}; \\
\quad \ident{attr}(\ident{attributeName}~:~\kvd{string},~\ident{value}~:~\kvd{any}):~\ident{JQuery}; \\
\}
\end{array}
\end{equation*}
The example demonstrates typical problems that arise when using type providers for language
interoperation. Not all TypeScript language constructs have direct equivalent in F\# and
the TypeScript type provider needs to map them to other constructs that are available:
\begin{itemize}
\item The interface file defines \ident{jQuery} as a global variable. F\# type providers cannot
  provide global bindings and so the variable is exposed as a static member of the imported type
  (\emph{i.e.}~\ident{j.jQuery}).

\item The interface \ident{JQueryStatic} specifies that the \ident{jQuery} object is callable
  (other members are omitted). This denotes that the corresponding JavaScript object is a
  callable function with other members. F\# does not allow ``calling an object'' and so
  this is mapped to a method \ident{j.jQuery.Invoke}.
\end{itemize}

\noindent
It is also worth noting that the \ident{attr} member is overloaded and uses an optional parameter.
However, both of these features are available in F\# and can be directly mapped. The following
snippet shows the runtime code that is generated when the types are erased:

\begin{equation*}
\begin{array}{l}
 \sem{~\ident{j.jQuery.Invoke}(\ident{command})~} ~= \\
 \qquad \kvd{let}~\ident{jq}~=~\ident{Emit.PropertyGetImpl}(\kvd{true}, \str{jQuery}, \lbrack|~|\rbrack) \\
 \qquad \ident{Emit.CallImpl}(\kvd{false}, \str{}, \lbrack|~ \ident{jq};~ \ident{command}~|\rbrack~) \\[0.75em]
 \sem{~\ident{jQuery}(\str{<input>})\ident{.attr}(\str{type},\str{checkbox})~} ~= \\
 \qquad \ident{Emit.CallImpl}(\kvd{false}, \str{attr}, \lbrack|~ \ident{jQuery}(\str{<input>});~\str{type};~\str{checkbox}~|\rbrack~)
\end{array}
\end{equation*}
The provided code is never actually \emph{executed}. As discussed in Section~\ref{sec:js-meta}, it
is translated to JavaScript (which then runs in the web browser). So, the above code can be seen
more as instructions for the translator. The mapping is straightforward -- member calls are
translated to \ident{Emit.CallImpl} (empty name denotes that the object itself is called) and
property getters are translated to \ident{Emit.PropertyGetImpl}. The first argument
denotes whether the call is static and the last argument is an array of arguments.

\subsubsection{Safety of cross-language type providers}
TypeScript does not have the traditional type soundness property \cite{ms-safets},
because of covariant generics. However, this is a property of \emph{running} TypeScript code, while
we use TypeScript only as a source of annotations for JavaScript libraries. Still, there is only a
weak guarantee that the library will adhere to the specification. We return to this issue
when discussing translation to JavaScript (Section~\ref{sec:js-meta}).

More importantly, we do not import the typing rules associated with TypeScript.
The type provider mechanism can only provide F\# types that then behave according to the
F\# typing rules. This means that we can import generics\footnote{Due to technical
limitations, this is not currently allowed by F\# type providers, but there are no
fundamental reasons for this.}, but this will not automatically allow using them
covariantly. In some cases, the F\# type system is simply more strict -- and in that case,
users have to use unsafe operations, such as $\ident{unbox}\langl \ident{bool} \rangl$
in the \ident{render} function. We return to the topic of unsafe operations in
Section~\ref{sec:js-meta}.

The above discussion highlights a broader point about using type providers for language
interoperability. For any language, its type system may be weaker or stricter in some ways:

\begin{itemize}
\item When interoperating with a weaker system, the type provider may need
  to map more types to a general type like \ident{object}. This makes the provided operations hard
  to use (as the source language is more flexible than F\#). An alternative is to use explicit
  annotations (like \textcolor{strclr}{\ttfamily d.ts} files for JavaScript).

\item If the imported language has a more precise type system than F\#, the type
  provider has to drop some of the information. This can be done safely for types in
  contravariant positions. In the unsafe case, the runtime needs to perform dynamic checks
  (Section~\ref{sec:conc-interop}).
\end{itemize}

\noindent
In case of TypeScript, we encounter both cases. In many cases, the return type is not
statically known and is exposed as \kvd{any}, which we then map to F\# \ident{object}
(and the developer has to use \ident{unbox}). However, TypeScript also supports limited
form of dependent typing (overloading on constants \cite{ms-typescript-09}). This cannot
be expressed in the F\# type system and the developer has to choose the right provided overload.

\subsubsection{Discussion of alternatives}
There are two options for calling JavaScript libraries such as jQuery. We can use annotations
-- written either in F\# or imported into F\#, or we can let the user write inline JavaScript.
A different approach is to discourage the use of existing JavaScript libraries as discussed in
Section~\ref{sec:js-lib}.

\paragraph{Embedding inline JavaScript.}
In general, the host language should not know about the sytnax of the embedded language. This can
be hanlded via quasi-quotations \cite{haskell-templ}, or by embedding JavaScript as strings (perhaps
with syntax checking at a later stage). In FunScript, the \ident{number} function is defined as:
\begin{equation*}
\begin{array}{l}
 \lbrack\langl\ident{JSEmit}(\str{return \{0\}*1.0;})\rangl\rbrack \\
 \kvd{let}~\ident{number}~(\ignp:\ident{obj}) : \ident{float} ~=~\ident{failwith}~\str{JavaScript stub should not be called.}
\end{array}
\end{equation*}
The example uses .NET attributes, which are meta-data attached to a function. When FunScript finds
a call to such annotated function, it replaces the call with the specified JavaScript.
The approach provides no guarantees about the inline JavaScript code. In practice, this feature
is used only for minimal JavaScript runtime, but the approach could be improved using a
parameterized type provider such as $\ident{JSCode}\langl\str{return \{0\}*1.0;}\rangl$, which
would check that the static parameter is syntactically correct JavaScript code. So, type providers
could be used to make this approach safer.

\vspace{-1em}
\paragraph{Other ways of writing annotations.}
Many compilers to JavaScript provide a way for writing type signatures for JavaScript
libraries in the host language. For example, the following shows a declaration of the
jQuery \ident{attr} method from a mapping for js\_of\_ocaml \cite{js_ocmal} (definitions in
SMLtoJs \cite{js_sml} are simpler):
\begin{equation*}
\begin{array}{l}
 \kvd{class type}~\ident{jQuery}~=~\kvd{object} \\
 \quad \kvd{method}~\ident{attr}~:~\ident{js\_string t} ~\rightarrow~ \ident{js\_string t optdef meth} \\
 \quad \kvd{method}~\ident{attr\_set}~:~\ident{js\_string t} ~\rightarrow~ \ident{js\_string t} ~\rightarrow~ \ident{jQuery t meth} \\
 \kvd{end}
\end{array}
\end{equation*}
The js\_of\_ocaml project uses the OCaml object model for calling JavaScript libraries.
Compared with FunScript, it is more explicit in importing JavaScript -- the \ident{meth}
type denotes a JavaScript method; \ident{optdef} specifies that the result may be
undefined and \ident{js\_string} denotes a JavaScript string. Also, note that the second method
is called \ident{attr\_set}. This is a simple naming trick -- OCaml does not support overloading
and js\_of\_ocaml simply ignores anything after underscore.

Using a type provider is similar to writing a code generator that turns TypeScript
\textcolor{strclr}{\ttfamily d.ts} files into the above OCaml annotation. The main difference
is that type providers do not produce any artefacts
(generated files) and can import entire repositories lazily.

%
%

\section{Compiling to JavaScript}
\label{sec:js}

JavaScript has become the \emph{lingua franca} of the web and an increasing number of programming
languages provide a way of compiling to JavaScript. In F\#, the first project was F\# WebTools
\cite{fsharp-webtools} in 2007. A more recent and complex framework called Websharper \cite{websharper-guis}
is available with full commercial support. In this paper, we use FunScript which is a lightweight
library focused just on translating F\# to JavaScript.

A wide range of choices is available when targeting JavaScript. The compilation to JavaScript can be
implemented as compiler back-end or as a library (Section~\ref{sec:js-meta}) and there are different
approaches to base libraries (Section~\ref{sec:js-lib}) and asynchronous computing
(Section~\ref{sec:js-async}).

\subsection{Lightweight meta-programming with quotations}
\label{sec:js-meta}

The FunScript library is based on F\# quotations \cite{fsharp-metaprog}. This means that the code
is compiled as ordinary F\#, but the compiler also stores marked blocks of code as data.
In FunScript, we typically need to translate the entire source file. To allow this, we instruct the
compiler to store code as data for the whole module using the \ident{ReflectedDefinition} attribute.
This is done at the beginning of the file:

\begin{equation*}
\begin{array}{l}
 \lbrack\langl\ident{ReflectedDefinition}\rangl\rbrack \\
 \kvd{module}~\ident{Program}
 \\[0.5em]
 \kvd{open}~\ident{FunScript} \\
 \kvd{open}~\ident{FunScript.TypeScript} \\
 \kvd{open}~\ident{FSharp.Data}
\end{array}
\end{equation*}
When marked with \ident{ReflectedDefinition}, the F\# compiler stores the body of all functions and
methods in the marked module as \emph{quotations} that can be retrieved at run-time.
Compared to writing a full compiler back-end, the lightweight approach to meta-programming makes it easy to
write a translator from F\# to other languages. This has been used for compiling to SQL queries, compiling
F\# code to CUDA or for Freebase queries \cite{fsharp-data}. For example, explicit quotations such as
$\langl @~ ((\ident{shift input}~-\hspace{-0.2em}1) ~+~ \ident{input})/2 ~@\rangl$
have been used for translating F\# code to GPU \cite{accelerator}.

Using the lightweight meta-programming approach in FunScript would not be appropriate if we wanted to
compile arbitrary existing F\# source code to JavaScript without any modifications. For that, using a
compiler back-end is a better choice. However, for the task solved in our case study (compiling
newly written F\# code), the approach works well.

\subsection{Accessing F\# and JavaScript libraries}
\label{sec:js-lib}

The F\# code in our case study uses a number of libraries. This includes the F\# core library (for
example, the \ident{List.map} function), standard .NET libraries (iteration using the \kvd{for} loop
uses .NET \ident{IEnumerable} interface) and JavaScript libraries. Accessing these libraries in
FunScript follows the F\# WebTools \cite{fsharp-webtools}.

The range of possible approaches to libraries when translating to JavaScript has two extreme cases.
We can attempt to port all libraries of the source language (F\#, Haskell or even .NET), or we
can ignore standard libraries and instead provide access to all JavaScript libraries. FunScript
stands in the middle -- it provides access to \emph{some} F\# and .NET libraries,
but relies  on JavaScript for more advanced functionality.

\paragraph{Mapping standard .NET and F\# libraries.}
The F\# ecosystem relies on .NET, so taking an F\# library and translating it to JavaScript without
any modification would require translating any (compiled) .NET library. This is where F\# differs,
for example, from Haskell which has a closed ecosystem and the ghcjs project \cite{haskell-ghcjs}
is thus capable of translating most standard libraries.

An interesting aspect of the case study is that we use F\# Data \cite{fsharp-data} type providers,
which is a standard F\# library that has not been built specifically for FunScript. Thus it is worth
explaining how type providers and quotations interact.
When we write code using type providers that provide erased types (both World Bank and TypeScript),
then the erasure happens before a quotation is captured. When we write code with explicit (or
implicit) quotation containing a provided method (1), the actual
quotation (2) contains just calls to the underlying runtime library:
\begin{equation*}
\begin{array}{llr}
 \langl @~\ident{data.}\lident{Czech Republic}~@\rangl &\hspace{10em}& (1)\\
 \langl @~\ident{data.GetCountries}().\ident{GetCountry}(\str{CZE}) ~@\rangl  && (2)
\end{array}
\end{equation*}
As a result, the FunScript component only needs to provide a client-side implementation of the underlying
operations \ident{GetCountries}, \ident{GetCountry} and \ident{AsyncGetIndicator}. This is done in the
same way as mappings for standard .NET libraries. Under the cover, \ident{AsyncGetIndicator} invokes
an AJAX request to the World Bank API. We provide more details about asynchronous execution in
Section~\ref{sec:js-async}.

\paragraph{Accessing standard JavaScript libraries.}
As discussed earlier, the focus of this case study is on integrating multiple ecosystems -- using
F\# type providers for data access and JavaScript libraries for visualization. In this approach
a high degree of integration between F\# and JavaScript is necessary.

An important problem with accessing JavaScript libraries is deciding which types should be
mapped to ordinary F\# types and which types should be treated as opaque JavaScript types that
can only be manipulated by JavaScript operations. Consider the following excerpt from the code
in Section~\ref{sec:case-loading}:
\begin{equation*}
\begin{array}{lll}
 \kvd{let}~\ident{o}~=~\ident{h.HighchartsOptions}() \\
 \ident{o.chart} \leftarrow \ident{h.HighchartsChartOptions}(\ident{renderTo}=\str{plc}) \\
 \ident{o.title} \leftarrow \ident{h.HighchartsTitleOptions}(\ident{text}~=~\ident{head}) \\
 \ident{o.series} \leftarrow \lbrack| ~ |\rbrack \\
 (\ldots)
 \\[0.5em]
 \ident{opts.series.push}(\ident{h.HighchartsSeriesOptions}(\ident{data}, \ident{name}))
\end{array}
\end{equation*}
In ordinary F\# code, we would not initialize \ident{o.series} to an empty array before
accessing it and we would not expect to use a \ident{push} method to append an element --
.NET arrays are mutable, but not resizable. Here, the first is necessary. When created,
the \ident{series} property of \ident{HighchartsOptions} is \kvd{undefined}. The \ident{push}
operation is added to standard .NET array, but it cannot be implemented for ordinary F\# code.
In summary, the options for mapping JavaScript libraries to F\# and the choices made by FunScript
are:

\begin{itemize}
\item Numeric types including \ident{int} and \ident{float} are mapped to JavaScript numbers.
  However, the floating-point arithmetic in JavaScript differs from the one in F\# and so this changes
  the semantics. Compilation of integer operators requires additional work (as $1/2$ is not an
  integer division in JavaScript).

\item FunScript maps F\# arrays to JavaScript arrays. We can define extension methods
  such as \ident{push} that make sense for JavaScript arrays. This lets us use standard
  F\# modules and functions such as \ident{Array.map}. An alternative is to use a
  separate type -- js\_of\_ocaml uses a separate type \ident{js\_array}.

\item Due to its .NET heritage, there are many types in F\# that have \kvd{null} as a valid
  value. This means that there is a reasonable precedent for allowing \kvd{undefined} values
  on types imported from JavaScript (but not for types defined in F\#). In contrast,
  js\_of\_ocaml is more explicit and uses \ident{optdef} type to denote potentially undefined
  values.
\end{itemize}

\noindent
In summary, the approach used by F\# is to reuse as much of F\# and .NET as easily possible. This
makes writing and reusing code easier and lets developers use existing familiar libraries and
idioms. However, it means that the semantics of F\# running as JavaScript is not always strictly
the same as the semantics of F\# running as compiled code.

\paragraph{Client-side asynchronous computations.}
\label{sec:js-async}
In F\#, asynchronous workflows \cite{fsharp-async} serve two purposes. First, they provide a
way for running multiple tasks in parallel. Second, they make it possible to write long-running
non-blocking code without the use of explicit callbacks. In our case study, we only use the
latter aspect. When loading data for selected countries in a loop, we fetch data asynchronously
(using AJAX). Under the cover, JavaScript triggers a callback, which then resumes the
asynchronous workflow.

FunScript uses a simple cooperative model based on continuations (which is similar to how Links
\cite{web-links} compilation works). In ordinary F\#, asynchronous workflows can be started in a
number of ways: \ident{Async.RunSynchronously} starts the work and blocks until it completes;
\ident{Async.Start} starts the work in the background and \ident{Async.StartImmediate} starts the
work on the current thread and runs callbacks on the same thread. Only the last one
can be mapped to JavaScript (FunScript translator will fail at translation-time if the other two are used).
Our approach is again to reuse standard F\# constructs, but interpret them in a different way that
fits better with the new execution environment.

%

\section{Relativized safety and semantics}

If there is one key message that the reader should take from this paper, it is the idea that both
the type safety property and the langauge semantics can be relativized. That is, when we write code
that uses the safe ML subset of F\#, we get all the good ML properties. However, to be able to
interoperate with the open world and other ecosystems, the core language can be reinterpreted -- in
a practically useful way but with semantics and safety that is relative with respect to the new
environment.

\subsection{Relativized type safety}
\label{sec:tp-relative}

Type providers for information sources, such as the World Bank provider, weaken the usual notion
of type safety. In general, they require that certain assumptions about the world do not change
between the state of the world at compile-time and state of the world at run-time. The TypeScript
provider is different and it is more interesting to consider runtime behaviour as discussed
in Section~\ref{sec:conc-interop}.

For type providers that provide access to external information sources, we can formulate a
\emph{relativized} notion of type safety. A full formalization is beyond the scope of this
paper, but the following provides an outline of such a theorem.

\begin{theorem*}[Relativized type safety]
Assume $\pi_{\ident{~WorldBank}}$ is the mapping implemented by the World Bank type provider,
$w_0, w_1$ are models of a (read-only) world that can be seen as functions defined on country-indicator
pairs and $\langl e, w \rangl \rightarrow e'$ is a one-step reduction on expressions that has
access to the World Bank data modelled by $w$.

Then, if $e$ is well-typed using a compile-time model of the world $w_0$ and the model of the world
used at run-time $w_1$ contains all country indicator pairs that may be accessed by $e$, then $e$
can make a reduction step, \emph{i.e.} $\langl e, w_1 \rangl \rightarrow e'$ and the type of $e'$
is the same as the type of $e$.
\end{theorem*}

\noindent
The theorem has the usual structure of a type safety theorem. The reduction is defined on pairs
consisting of expression and a world $w$. The result of the reduction is only an expression and
so it cannot modify the world. Even though we give only a brief sketch, it demonstrates two
important points about relativized type safety that are relevant to many type providers.

First, we distinguish between the state of the world $w_0$ that is used at compile-time and the
state $w_1$ used at run-time. If these were the same, the additional assumption would always hold.
Second, we require that all countries and indicators that \emph{may be accessed} need to be
available. This is an over-approximation, but it does \emph{not} say that
the run-time world must contains \emph{all} countries and indicators.

\subsection{Relativized semantics}
\label{sec:js-relative}

When compiling any language to JavaScript, we can treat JavaScript as a low-level runtime
(and use the asm.js \cite{asm-js} subset for efficiency), or we can use it as
higher-level language and map many source language constructs to corresponding JavaScript
constructs. In this case study, we used the latter. This enables interop with JavaScript, but
it makes it hard (if at all possible) to preserve the original F\# semantics.

This is perhaps a controversial approach, but it fits well with F\#, because it is similar to
how F\# handles interop with .NET. When you do not use .NET objects, you do not need type
annotations and you do not have to handle \kvd{null} values. If you use .NET, you have to
accept those at some level. More generally, we could call this property \emph{relativized
semantics}:

\begin{quotation}
\noindent
\emph{When you use the ML subset of F\#, the program will behave in the same way regardless
of the environment and you are guaranteed the usual ML properties. In other environments,
the properties are not guaranteed.}
\end{quotation}

\noindent
Making the notion of \emph{relativized semantics} formally precise is outside of the scope
of this paper. An important aspect is specifying the boundary (see Section~\ref{sec:conc-interop}).
The problem is similar to writing monadic computations -- each monad is different (and has
different properties), but there are certain laws that always hold. For full cross-compilation,
we need a similar set of laws, but for all ML language constructs.


\section{Related and further work}
\label{sec:conc-interop}

\paragraph{Related work.}

There are a number of projects that solve similar problems to the ones described in this paper.
In the F\# ecosystem, the first project compiling to JavaScript was F\# WebTools \cite{fsharp-webtools},
which also supported asynchronous computations (although using a separate computation type)
and interop with JavaScript. More recently, WebSharper \cite{websharper-piglets,websharper-guis}
is a more complete framework for web development that also provides composable abstractions for
building forms (based on formlets \cite{links-formlets}) and entire applications.

Other statically-typed functional languages that have some way of running as JavaScript
include OCaml \cite{js_ocmal}, SML \cite{js_sml} and Haskell \cite{haskell-ghcjs}. Compared to
our work, all three are stricter in preserving the semantics of the source language, but
do not provide as smooth JavaScript integration.

\vspace{-1em}
\paragraph{Further work.}
The programming model used in our case study combines dynamically typed components (JavaScript
libraries) and statically typed components (code written in F\#). This suggests a relation with
the work on gradual typing \cite{gradual-fun,gradual-oop} and blame tracking \cite{blame-well}. However,
in our work, the distinction between the statically and dynamically typed parts (and the boundary)
is less explicit. However, it would be interesting to see if gradual typing and blame can be
used to formally define our informal notion of \emph{relativized semantics} introduced in
Section~\ref{sec:js-relative}.


\section{Conclusions}

The key idea of this paper is that writing modern applications for the web requires us to
reconsider many assumptions that functional language designers take for granted. The novelty
of this paper is not in any single technology it presents, but in the combination it shows.
For this reason, we used the form of a case study, using a significant example to guide our
discussion.

There are two main assumptions that we reconsider. The first assumption is that programs
operate in a closed world. Instead, modern application access a range of external services
and information-sources. In our case study, these are accessed using type providers.
The second assumption is that we can fix the runtime semantics. This becomes difficult when
the same code is compiled for diverse execution environments such as .NET, JavaScript or
even CUDA.

If there is one thing that the reader should remember from this case study, it is the
idea about F\# saying that ``\emph{when you use it as ML, it behaves as ML}''. That is, when
you use the ML subset of F\# and do not access external services, information or .NET and
JavaScript libraries, you still get all the good properties of ML. However, when you access
external information and run your code as JavaScript, you get a weaker notion of safety that
we called \emph{relativized type safety} and a weaker runtime guarantees that we called
\emph{relativized semantics}. Nevertheless, the case study shows that these are enough to
let us benefit from the functional-first statically-typed programming paradigm in the age of web.
\newpage


\nocite{*}
\bibliographystyle{eptcs}
\bibliography{generic}

\begin{thebibliography}{10}
\providecommand{\bibitemdeclare}[2]{}
\providecommand{\surnamestart}{}
\providecommand{\surnameend}{}
\providecommand{\urlprefix}{Available at }
\providecommand{\url}[1]{\texttt{#1}}
\providecommand{\href}[2]{\texttt{#2}}
\providecommand{\urlalt}[2]{\href{#1}{#2}}
\providecommand{\doi}[1]{doi:\urlalt{http://dx.doi.org/#1}{#1}}
\providecommand{\bibinfo}[2]{#2}

\bibitemdeclare{}{data-wb-schter}
\bibitem{data-wb-schter}
\bibinfo{author}{The~World \surnamestart Bank\surnameend}
  (\bibinfo{year}{2015}): \emph{\bibinfo{title}{School enrollment, tertiary (\%
  gross)}}.
\newblock \urlprefix\url{http://data.worldbank.org/indicator/SE.TER.ENRR}.

\bibitemdeclare{inproceedings}{websharper-guis}
\bibitem{websharper-guis}
\bibinfo{author}{Joel \surnamestart Bjornson\surnameend},
  \bibinfo{author}{Anton \surnamestart Tayanovskyy\surnameend} \&
  \bibinfo{author}{Adam \surnamestart Granicz\surnameend}
  (\bibinfo{year}{2011}): \emph{\bibinfo{title}{Composing Reactive GUIs in F\#
  Using WebSharper}}.
\newblock In: {\sl \bibinfo{booktitle}{Proceedings of IFL}},
  \bibinfo{series}{IFL'10}, pp. \bibinfo{pages}{203--216},
  \doi{10.1007/978-3-642-24276-2\_13}.

\bibitemdeclare{}{fsharp-rprovider}
\bibitem{fsharp-rprovider}
\bibinfo{author}{\surnamestart {BlueMountain Capital and
  Contributors}\surnameend} (\bibinfo{year}{2015}): \emph{\bibinfo{title}{F\# R
  Type Provider}}.
\newblock \urlprefix\url{http://bluemountaincapital.github.io/FSharpRProvider}.

\bibitemdeclare{}{fsharp-funscript}
\bibitem{fsharp-funscript}
\bibinfo{author}{Zach \surnamestart Bray\surnameend} \&
  \bibinfo{author}{\surnamestart Contributors\surnameend}
  (\bibinfo{year}{2015}): \emph{\bibinfo{title}{FunScript: F\# to JavaScript
  with type providers}}.
\newblock \urlprefix\url{http://funscript.info}.

\bibitemdeclare{book}{philosophy-thing}
\bibitem{philosophy-thing}
\bibinfo{author}{Alan~F \surnamestart Chalmers\surnameend}
  (\bibinfo{year}{2013}): \emph{\bibinfo{title}{What is this thing called
  science?}}
\newblock \bibinfo{publisher}{Open University Press}, \doi{10.1007/BF00174905}.

\bibitemdeclare{inproceedings}{web-links}
\bibitem{web-links}
\bibinfo{author}{Ezra \surnamestart Cooper\surnameend}, \bibinfo{author}{Sam
  \surnamestart Lindley\surnameend}, \bibinfo{author}{Philip \surnamestart
  Wadler\surnameend} \& \bibinfo{author}{Jeremy \surnamestart
  Yallop\surnameend} (\bibinfo{year}{2007}): \emph{\bibinfo{title}{Links: Web
  Programming Without Tiers}}.
\newblock In: {\sl \bibinfo{booktitle}{Proceedings of FMCO}},
  \bibinfo{series}{FMCO'06}, pp. \bibinfo{pages}{266--296},
  \doi{10.1007/978-3-540-74792-5\_12}.
\newblock \urlprefix\url{http://dl.acm.org/citation.cfm?id=1777707.1777724}.

\bibitemdeclare{inproceedings}{links-formlets}
\bibitem{links-formlets}
\bibinfo{author}{Ezra \surnamestart Cooper\surnameend}, \bibinfo{author}{Sam
  \surnamestart Lindley\surnameend}, \bibinfo{author}{Philip \surnamestart
  Wadler\surnameend} \& \bibinfo{author}{Jeremy \surnamestart
  Yallop\surnameend} (\bibinfo{year}{2008}): \emph{\bibinfo{title}{The Essence
  of Form Abstraction}}.
\newblock In: {\sl \bibinfo{booktitle}{Proceedings of APLAS}},
  \bibinfo{series}{APLAS '08}, pp. \bibinfo{pages}{205--220},
  \doi{10.1007/978-3-540-89330-1\_15}.

\bibitemdeclare{}{asm-js}
\bibitem{asm-js}
\bibinfo{author}{Alon~Zakai \surnamestart David~Herman\surnameend, Luke~Wagner}
  (\bibinfo{year}{2014}): \emph{\bibinfo{title}{asm.js -- Working Draft -- 18
  August 2014}}.
\newblock \urlprefix\url{http://asmjs.org/spec/latest/}.

\bibitemdeclare{inproceedings}{websharper-piglets}
\bibitem{websharper-piglets}
\bibinfo{author}{Lo\"{\i}c \surnamestart Denuzi\`{e}re\surnameend},
  \bibinfo{author}{Ernesto \surnamestart Rodriguez\surnameend} \&
  \bibinfo{author}{Adam \surnamestart Granicz\surnameend}
  (\bibinfo{year}{2014}): \emph{\bibinfo{title}{Piglets to the Rescue:
  Declarative User Interface Specification with Pluggable View Models}}.
\newblock In: {\sl \bibinfo{booktitle}{Proceedings of IFL}},
  \bibinfo{series}{IFL '13}, \bibinfo{publisher}{ACM}, pp.
  \bibinfo{pages}{105:105--105:115}, \doi{10.1145/2620678.2620689}.

\bibitemdeclare{inproceedings}{js_sml}
\bibitem{js_sml}
\bibinfo{author}{Martin \surnamestart Elsman\surnameend}
  (\bibinfo{year}{2011}): \emph{\bibinfo{title}{SMLtoJs: Hosting a Standard ML
  Compiler in a Web Browser}}.
\newblock In: {\sl \bibinfo{booktitle}{Proceedings of PLASTIC}},
  \bibinfo{series}{PLASTIC '11}, \bibinfo{publisher}{ACM}, pp.
  \bibinfo{pages}{39--48}, \doi{10.1145/2093328.2093336}.

\bibitemdeclare{inproceedings}{haskell-ffi}
\bibitem{haskell-ffi}
\bibinfo{author}{Sigbjorn \surnamestart Finne\surnameend},
  \bibinfo{author}{Daan \surnamestart Leijen\surnameend}, \bibinfo{author}{Erik
  \surnamestart Meijer\surnameend} \& \bibinfo{author}{Simon \surnamestart
  Peyton~Jones\surnameend} (\bibinfo{year}{1999}):
  \emph{\bibinfo{title}{Calling Hell from Heaven and Heaven from Hell}}.
\newblock In: {\sl \bibinfo{booktitle}{Proceedings of ICFP}},
  \bibinfo{series}{ICFP '99}, \bibinfo{publisher}{ACM}, pp.
  \bibinfo{pages}{114--125}, \doi{10.1145/317636.317790}.

\bibitemdeclare{}{google-freebase}
\bibitem{google-freebase}
\bibinfo{author}{\surnamestart Google\surnameend} (\bibinfo{year}{2015}):
  \emph{\bibinfo{title}{Freebase: A community-curated database of well-known
  people, places, and things}}.
\newblock \urlprefix\url{http://www.freebase.com}.

\bibitemdeclare{book}{dj-handbook}
\bibitem{dj-handbook}
\bibinfo{author}{Jonathan \surnamestart Gray\surnameend}, \bibinfo{author}{Lucy
  \surnamestart Chambers\surnameend} \& \bibinfo{author}{Liliana \surnamestart
  Bounegru\surnameend} (\bibinfo{year}{2012}): \emph{\bibinfo{title}{The data
  journalism handbook}}.
\newblock \bibinfo{publisher}{O'Reilly Media}.

\bibitemdeclare{}{fsharp-fcs}
\bibitem{fsharp-fcs}
\bibinfo{author}{The F\# Core~Engineering \surnamestart Group\surnameend}
  (\bibinfo{year}{2015}): \emph{\bibinfo{title}{F\# Compiler Services: Editor
  services}}.
\newblock
  \urlprefix\url{http://fsharp.github.io/FSharp.Compiler.Service/editor.html}.

\bibitemdeclare{incollection}{philosophy-lakatos}
\bibitem{philosophy-lakatos}
\bibinfo{author}{Imre \surnamestart Lakatos\surnameend} (\bibinfo{year}{1970}):
  \emph{\bibinfo{title}{Falsification and the Methodology of Scientific
  Research Programmes}}.
\newblock In \bibinfo{editor}{Imre \surnamestart Lakatos\surnameend} \&
  \bibinfo{editor}{Alan \surnamestart Musgrave\surnameend}, editors: {\sl
  \bibinfo{booktitle}{Criticism and the Growth of Knowledge}},
  \bibinfo{publisher}{Cambridge University Press}, pp.
  \bibinfo{pages}{91--196}, \doi{10.1007/978-94-010-1863-0\_14}.

\bibitemdeclare{}{haskell-ghcjs}
\bibitem{haskell-ghcjs}
\bibinfo{author}{Hamish~Mackenzie \surnamestart Luite~Stegeman\surnameend} \&
  \bibinfo{author}{\surnamestart Contributors\surnameend}
  (\bibinfo{year}{2015}): \emph{\bibinfo{title}{{ghcjs: Project homepage}}}.
\newblock \urlprefix\url{https://github.com/ghcjs}.

\bibitemdeclare{}{ms-typescript}
\bibitem{ms-typescript}
\bibinfo{author}{\surnamestart Microsoft\surnameend} \&
  \bibinfo{author}{\surnamestart Contributors\surnameend}
  (\bibinfo{year}{2015}): \emph{\bibinfo{title}{TypeScript}}.
\newblock \urlprefix\url{http://typescriptlang.org}.

\bibitemdeclare{}{fsharp-paramsdict}
\bibitem{fsharp-paramsdict}
\bibinfo{author}{Tomas \surnamestart Petricek\surnameend}
  (\bibinfo{year}{2014}): \emph{\bibinfo{title}{F\# Language: Allow ``params''
  dictionaries as method arguments}}.
\newblock
  \urlprefix\url{https://fslang.uservoice.com/forums/245727/suggestions/5975840}.

\bibitemdeclare{inproceedings}{philosophy-pl}
\bibitem{philosophy-pl}
\bibinfo{author}{Tomas \surnamestart Petricek\surnameend}
  (\bibinfo{year}{2014}): \emph{\bibinfo{title}{What can Programming Language
  Research Learn from the Philosophy of Science?}}
\newblock In: {\sl \bibinfo{booktitle}{Proceedings of the 50th Anniversary
  Convention of the AISB}}.

\bibitemdeclare{}{fsharp-data}
\bibitem{fsharp-data}
\bibinfo{author}{Tomas \surnamestart Petricek\surnameend},
  \bibinfo{author}{Gustavo \surnamestart Guerra\surnameend} \&
  \bibinfo{author}{\surnamestart Contributors\surnameend}
  (\bibinfo{year}{2015}): \emph{\bibinfo{title}{F\# Data: Library for Data
  Access}}.
\newblock \urlprefix\url{http://fsharp.github.io/FSharp.Data/}.

\bibitemdeclare{}{fsharp-webtools}
\bibitem{fsharp-webtools}
\bibinfo{author}{Tomas \surnamestart Petricek\surnameend} \&
  \bibinfo{author}{Don \surnamestart Syme\surnameend} (\bibinfo{year}{2007}):
  \emph{\bibinfo{title}{F\# Web Tools: Rich client/server web applications in
  F\#}}.
\newblock \bibinfo{howpublished}{Unpublished draft, submitted to ML Workshop
  2007}.
\newblock \urlprefix\url{http://tomasp.net/academic/articles/fswebtools}.

\bibitemdeclare{inproceedings}{fsharp-zoo}
\bibitem{fsharp-zoo}
\bibinfo{author}{Tomas \surnamestart Petricek\surnameend} \&
  \bibinfo{author}{Don \surnamestart Syme\surnameend} (\bibinfo{year}{2014}):
  \emph{\bibinfo{title}{The F\# Computation Expression Zoo}}.
\newblock In: {\sl \bibinfo{booktitle}{Proceedings of PADL}},
  \bibinfo{series}{PADL 2014}, \bibinfo{publisher}{Springer-Verlag New York,
  Inc.}, pp. \bibinfo{pages}{33--48}, \doi{10.1007/978-3-319-04132-2\_3}.

\bibitemdeclare{inproceedings}{haskell-imperative}
\bibitem{haskell-imperative}
\bibinfo{author}{Simon~L. \surnamestart Peyton~Jones\surnameend} \&
  \bibinfo{author}{Philip \surnamestart Wadler\surnameend}
  (\bibinfo{year}{1993}): \emph{\bibinfo{title}{Imperative Functional
  Programming}}.
\newblock In: {\sl \bibinfo{booktitle}{Proceedings of POPL}},
  \bibinfo{series}{POPL '93}, \bibinfo{publisher}{ACM}, pp.
  \bibinfo{pages}{71--84}, \doi{10.1145/158511.158524}.

\bibitemdeclare{techreport}{ms-safets}
\bibitem{ms-safets}
\bibinfo{author}{Aseem \surnamestart Rastogi\surnameend},
  \bibinfo{author}{Nikhil \surnamestart Swamy\surnameend},
  \bibinfo{author}{Cedric \surnamestart Fournet\surnameend},
  \bibinfo{author}{Gavin \surnamestart Bierman\surnameend} \&
  \bibinfo{author}{Panagiotis \surnamestart Vekris\surnameend}
  (\bibinfo{year}{2014}): \emph{\bibinfo{title}{Safe \& Efficient Gradual
  Typing for TypeScript}}.
\newblock \bibinfo{type}{Technical Report} \bibinfo{number}{MSR-TR-2014-99},
  \bibinfo{institution}{Microsoft Research}, \doi{10.1145/2676726.2676971}.
\newblock \urlprefix\url{http://research.microsoft.com/apps/pubs/?id=224900}.

\bibitemdeclare{inproceedings}{haskell-templ}
\bibitem{haskell-templ}
\bibinfo{author}{Tim \surnamestart Sheard\surnameend} \&
  \bibinfo{author}{Simon~Peyton \surnamestart Jones\surnameend}
  (\bibinfo{year}{2002}): \emph{\bibinfo{title}{Template meta-programming for
  Haskell}}.
\newblock In: {\sl \bibinfo{booktitle}{Proceedings of the 2002 ACM SIGPLAN
  workshop on Haskell}}, \bibinfo{organization}{ACM}, pp.
  \bibinfo{pages}{1--16}, \doi{10.1145/636517.636528}.

\bibitemdeclare{incollection}{gradual-oop}
\bibitem{gradual-oop}
\bibinfo{author}{Jeremy \surnamestart Siek\surnameend} \&
  \bibinfo{author}{Walid \surnamestart Taha\surnameend} (\bibinfo{year}{2007}):
  \emph{\bibinfo{title}{Gradual typing for objects}}.
\newblock In: {\sl \bibinfo{booktitle}{ECOOP 2007--Object-Oriented
  Programming}}, \bibinfo{publisher}{Springer}, pp. \bibinfo{pages}{2--27},
  \doi{10.1007/978-3-540-73589-2\_2}.

\bibitemdeclare{inproceedings}{gradual-fun}
\bibitem{gradual-fun}
\bibinfo{author}{Jeremy~G \surnamestart Siek\surnameend} \&
  \bibinfo{author}{Walid \surnamestart Taha\surnameend} (\bibinfo{year}{2006}):
  \emph{\bibinfo{title}{Gradual typing for functional languages}}.
\newblock In: {\sl \bibinfo{booktitle}{Scheme and Functional Programming
  Workshop}}, \bibinfo{volume}{6}, pp. \bibinfo{pages}{81--92},
  \doi{10.1.1.61.8890}.

\bibitemdeclare{inproceedings}{accelerator}
\bibitem{accelerator}
\bibinfo{author}{Satnam \surnamestart Singh\surnameend} (\bibinfo{year}{2010}):
  \emph{\bibinfo{title}{Declarative Data-parallel Programming with the
  Accelerator System}}.
\newblock In: {\sl \bibinfo{booktitle}{Proceedings of DAMP}},
  \bibinfo{series}{DAMP '10}, \bibinfo{publisher}{ACM}, pp.
  \bibinfo{pages}{1--2}, \doi{10.1145/1708046.1708048}.

\bibitemdeclare{inproceedings}{fsharp-metaprog}
\bibitem{fsharp-metaprog}
\bibinfo{author}{Don \surnamestart Syme\surnameend} (\bibinfo{year}{2006}):
  \emph{\bibinfo{title}{Leveraging .NET Meta-programming Components from F\#:
  Integrated Queries and Interoperable Heterogeneous Execution}}.
\newblock In: {\sl \bibinfo{booktitle}{Proceedings of ML Workshop}},
  \bibinfo{series}{ML '06}, \bibinfo{publisher}{ACM}, pp.
  \bibinfo{pages}{43--54}, \doi{10.1145/1159876.1159884}.

\bibitemdeclare{techreport}{fsharp-typeprov}
\bibitem{fsharp-typeprov}
\bibinfo{author}{Don \surnamestart Syme\surnameend},
  \bibinfo{author}{K.~\surnamestart Battocchi\surnameend},
  \bibinfo{author}{K.~\surnamestart Takeda\surnameend},
  \bibinfo{author}{D.~\surnamestart Malayeri\surnameend},
  \bibinfo{author}{J.~\surnamestart Fisher\surnameend},
  \bibinfo{author}{J.~\surnamestart Hu\surnameend},
  \bibinfo{author}{T.~\surnamestart Liu\surnameend},
  \bibinfo{author}{B.~\surnamestart McNamara\surnameend},
  \bibinfo{author}{D.~\surnamestart Quirk\surnameend},
  \bibinfo{author}{M.~\surnamestart Taveggia\surnameend},
  \bibinfo{author}{W.~\surnamestart Chae\surnameend},
  \bibinfo{author}{U.~\surnamestart Matsveyeu\surnameend} \&
  \bibinfo{author}{T.~\surnamestart Petricek\surnameend}
  (\bibinfo{year}{2012}): \emph{\bibinfo{title}{Strongly-typed language support
  for internet-scale information sources}}.
\newblock \bibinfo{type}{Technical Report} \bibinfo{number}{MSR-TR-2012-101},
  \bibinfo{institution}{Microsoft Research}.
\newblock \urlprefix\url{http://research.microsoft.com/apps/pubs/?id=173076}.

\bibitemdeclare{inproceedings}{fsharp-async}
\bibitem{fsharp-async}
\bibinfo{author}{Don \surnamestart Syme\surnameend}, \bibinfo{author}{Tomas
  \surnamestart Petricek\surnameend} \& \bibinfo{author}{Dmitry \surnamestart
  Lomov\surnameend} (\bibinfo{year}{2011}): \emph{\bibinfo{title}{The F\#
  Asynchronous Programming Model}}.
\newblock In: {\sl \bibinfo{booktitle}{Proceedings of PADL}},
  \bibinfo{series}{PADL'11}, pp. \bibinfo{pages}{175--189},
  \doi{10.1007/978-3-642-18378-2\_15}.

\bibitemdeclare{}{ms-typescript-09}
\bibitem{ms-typescript-09}
\bibinfo{author}{Jonathan \surnamestart Turner\surnameend}
  (\bibinfo{year}{2013}): \emph{\bibinfo{title}{Announcing TypeScript 0.9}}.
\newblock
  \urlprefix\url{http://blogs.msdn.com/b/typescript/archive/2013/06/18/announcing-typescript-0-9.aspx}.

\bibitemdeclare{article}{js_ocmal}
\bibitem{js_ocmal}
\bibinfo{author}{J{\'e}r{\^o}me \surnamestart Vouillon\surnameend} \&
  \bibinfo{author}{Vincent \surnamestart Balat\surnameend}
  (\bibinfo{year}{2013}): \emph{\bibinfo{title}{From bytecode to javascript:
  the js\_of\_ocaml compiler}}.
\newblock {\sl \bibinfo{journal}{Software: Practice and Experience}},
  \doi{10.1.1.224.7457}.

\bibitemdeclare{inproceedings}{blame-well}
\bibitem{blame-well}
\bibinfo{author}{Philip \surnamestart Wadler\surnameend} \&
  \bibinfo{author}{Robert~Bruce \surnamestart Findler\surnameend}
  (\bibinfo{year}{2009}): \emph{\bibinfo{title}{Well-Typed Programs Can'T Be
  Blamed}}.
\newblock In: {\sl \bibinfo{booktitle}{Proceedings of ESOP}},
  \bibinfo{series}{ESOP '09}, pp. \bibinfo{pages}{1--16},
  \doi{10.1007/978-3-642-00590-9\_1}.

\end{thebibliography}
\end{document}